\documentclass[pra,tightenlines,a4paper,preprint,amsmath,aps]{revtex4}
\usepackage{dcolumn,bm}

\begin{document}

\title{\small Estimation of characteristic size of ferromagnetic clusters forming above $T_C$ in Nd$_{0.75}$Ba$_{0.25}$MnO$_3$ manganite}
\author{A.V.~Lazuta, V.A.~Ryzhov, and V.V.~Runov}
\address{Petersburg Nuclear Physics Institute of RAS, Gatchina,
St. Petersburg, 188300, Russia}
\author{I.O.~Troaynchuk}
\address{Institute of Physics of Solids and Semiconductors,
National Academy of Sciences, ul. P. Brovki 17, Minsk, 220072,
Belarus}
\date{recieved; accepted}

\begin{abstract}
We present the data on depolarization of polarized neutron beam and second harmonic of magnetization ($M_2$) for Nd$_{1-x}$Ba$_x$MnO$_3$ (x =  0.23, the Curie temperature $T_C \approx$ 124 K; x = 0.25,  $T_C \approx$ 129 K) manganites. The depolarization starts to develop below $T\mbox{*} \approx$ 147 K $> T_C$ for both samples, being larger in x = 0.25 compound. This evidences the arising of a ferromagnetic (F) cluster phase below $T$* and a growth of its relative volume fraction with increasing doping concentration that agrees with the previously published results of $M_2$ study. A characteristic size of the F clusters and their concentration are estimated combining the neutron depolarization and $M_2$ data for x = 0.25 manganite.
\end{abstract}

\maketitle


\newpage



The interest in the study of the doped perovskite manganites is due to their unusual magnetic and electronic properties, some aspects of which are not well understood. An important problem is the origination of an inhomogeneous magnetic state above Curie temperature, $T_C$, in these compounds [1]. 

Nd$_{1-x}$Ba$_x$MnO$_3$ is a series of doped manganites which exhibits a transition from paramagnetic (P) to ferromagnetic (F) state for 0.2 $\leq x \leq$ 0.35 ($T_C \sim$ 120 K) [2]. For $x \geq x_{IM} \approx$ 0.3, these compounds show metallic behavior below $T_C$, whereas for $x <$ 0.3 they remain the insulators in the magnetic ordered phase. The arising of the inhomogeneous magnetic state above $T_C$ in process of a development of the second order transition was observed in the insulating Nd$_{1-x}$Ba$_x$MnO$_3$ ($x$ = 0.23, 0.25) single crystals. It is characterized by the appearance of a magnetic phase with the strong nonlinear properties in the weak magnetic fields below $T\mbox{*} (\approx T_C$ + 20 K) [3-7]. The unconventional behavior was attributed to the F metallic regions which originate in the P matrix. The F clusters as well as the unconventional behavior above $T_C$ were found in the other doped manganites [8-13]. This suggests that this phenomenon is an intrinsic property of these compounds. At the same time, the properties of this clustered state are not well elucidated, specifically a characteristic size of the F clusters is under question. 

The data on the magnetic, structural and transport properties for $x$ = 0.23 and 0.25 NdBa manganites were published earlier [3-7]. According to the second harmonic of magnetization ($M_2$) measurements, the signal arising below $T$* from the F metallic regions indicated an increasing in a volume fraction of the F phase at $T \rightarrow T_C$, this fraction being larger in the $x$ = 0.25 compound. 
 
 In this paper the data on a depolarization of a transmitted polarized neutron beam are presented for both manganites. Also the additional data on the field dependence of the $M_2$ at $T$ =142.6 K for $x$ = 0.25 crystal are reported. The combined data on the depolarization and on $M_2(H)$ for this compound are used in the quantitative analysis. It shows that the single-domain F regions, which above $T_C$ are in a close to superparamagnetic (SPM) regime, can reveal the strong nonlinear behavior in the weak magnetic fields. The analysis allows us to estimate the parameters of these F clusters (a magnetization, a characteristic size, a concentration, a magnetic anisotropy and a relaxation rate of the magnetization). 
 
The same Nd$_{1-x}$Ba$_x$MnO$_3$ ($x$ = 0.25) single crystal as in Refs. (6,7) was employed at the $M_2$ measurements. It was grown by the flux melt technique, using the BaO-B$_2$O$_3$-BaF$_2$ ternary system as a solvent [14].  Structural single phase of the crystal was confirmed by the X-ray and neutron diffractions. The cation composition of the crystal was determined by X-ray fluorescent analysis. The average oxidation state of manganese determined by photometry [15] was found to be close to the value expected for the stoichiometric oxygen content. The powder samples used in the depolarization measurements were prepared by the traditional manner [2]. Structural single phase of the samples was controlled by X-ray diffraction, and the oxygen content was determined by a thermogravimetric analysis. 

The polarization measurements were performed using a small-angle polarized neutron setup ``Vector'' with a wavelength $\lambda$ = 9.2 \AA, $\triangle \lambda/\lambda$ = 0.25 (WWR-M reactor Gatchina) [16]. 

The additional measurements of the second harmonic of the longitudinal component of magnetization $M_2$ were performed in the parallel steady and alternating magnetic fields $H + h$sin$\omega t (h \leq$ 37 Oe,  $\omega/2\pi \approx$ 15.7 MHz) in the temperature range $T* \geq T \geq T_C$. The Re$M_2$ and Im$M_2$ parts of $M_2$ were simultaneously recorded as the functions of $H$. This field was scanned symmetrically relative to the point $H$=0 for detecting a field hysteresis of the signal. The amplitude of $H$-scan was 300 Oe. A condition $M_2 \propto h^2$ was satisfied in the measurements. An installation and a method of separation of the $M_2$-phase components have been described previously [17]. 

Fig.~1 shows the temperature dependence of the polarization for Nd$_{1-x}$Ba$_x$MnO$_3 \; x$ = 0.23, 0.25 polycrystalline manganites. It is seen that in both samples the depolarization appears just below $T\mbox{*} \approx$ 147 K $> T_C \, (T_C \approx$ 124 K for $x$ =0.23 and 129 K for $x$ =0.25), indicating the origination of the F regions, and it increases at decreasing temperature. The depolarization is larger for $x$ = 0.25 sample. These results confirm the $M_2(T, H)$ data [4,6].

The $M_2$ response of the F clusters coexists with the critical contribution of the paramagnetic matrix, which also increases with decreasing temperature. Therefore, it is convenient for the analysis to choose the $M_2(H,T)$ curves of the $x$ = 0.25 compound at $T$ = 142.6 K (well above $T_C$), which are displayed in Fig.~2. A characteristic kink in the Re$M_2(H)$ dependence, which is due to a competition of the paramagnetic critical contribution and the signal of the F regions, is clearly seen at this temperature. The Im$M_2(H)$ is mainly determined by the F clusters [6]. Additionally, the critical paramagnetic contribution is proportional to $H$ (see below) and the demagnetization corrections are small here. The subsystem of F clusters can be considered as an ensemble of single domain F particles in nearly superparamagnetic regime since the field hysteresis of $M_2(H)$ dependences is small [4,6]. Nevertheless, the presence of this small $H$-hysteresis can be seen in Im$M_2(H)$ component (Fig.~2b) as an incomplete inversion symmetry of the curve relative to the point $H$ = 0.

Let us go to the analysis. The neutron polarization $P$ after passing of the sample with the ferromagnetic regions of a small concentration is given by [18]
\begin{equation}
P = P_0 exp[-4/3(\gamma_n B/\nu)^{2}R \widetilde{C}^{1/3} L],
\end{equation}


\noindent
where $P_0$ is the initial polarization directed along the beam, $\gamma_n$ is the gyromagnetic ratio of the neutron, $\nu \approx$ 4.3$\cdot 10^4$ cm/s is the neutron velocity, $B = 4\pi g\mu_B<S>/V_0$ is the induction in the F clusters, $<S>$ is the value of the spin at temperature $T, V_0$  is  the volume per magnetic atom (58.6 \AA $^3$ [7]), $R$ is the mean radius of the F regions, $\tilde{C}$  is the relative concentration of the superparamagnetic phase and $L \approx$ 0.2 cm is the thickness of the sample. We have three unknown parameters $B, R, \widetilde{C}$  in (1) and need the additional $M_2$ results to find them from experimental data. 

Going over the $M_2$ analysis, let us consider a possible magnetic anisotropy of the F regions. Our compound has the {\it Pbnm} space group with a relationship between the lattice parameters $a > b \approx c/\sqrt{2}$ [7] that suggests a dominating uniaxial anisotropy directed along the $a$-axis. We assume an easy axis character of the anisotropy. This assumption is supported by our preliminarily magnetic resonance measurements in the $x$ = 0.3 single crystal (with a close to the $x$ = 0.25 manganite crystal structure) where the resonance signal from the F regions is observed above $T_C$. Such a signal is not detected at $x$ = 0.25 [7] because of a small concentration of the F clusters. As a result, the Hamiltonian of the single-domain F region is given by    
\begin{equation}
H = - N[\vec {\mu} \vec H + K(\vec e \vec {e}_{\mu})^2], 
\end{equation}


\noindent
where $N$ is the number of the spins in the region, $\mu = g\mu_B <S>, g \approx$ 2  is the  $g$ - factor, $H$ is the steady magnetic field, $K$ is the effective uniaxial anisotropy per a magnetic atom which can include a shape anisotropy, $\vec e$  is the direction of the uniaxial anisotropy and $\vec {e}_{\mu} = \vec {\mu}/\mu$. The $M_2(H)$ data, which obtained for the magnetic fields directed along the different crystal axes, do not reveal an orientation dependence. It suggests a twin structure with a nearly equal population of the structural domains. We will show below that the domains with easy axis give the main contribution to $M_2(H)$ for $\vec H$ directed along the crystallographic axes. In this case, the longitudinal response of the second order for single-domain F cluster can be written as [4]: 
\begin{equation}
\widetilde {M}_2 (H, \omega)/h^2 = \frac{\Gamma(H)}{-2i\omega + \Gamma (H)} \chi_2 (H) - \frac{i\omega}{\Gamma (H)} \frac{\partial \Gamma (H)/\partial H}{(-2i\omega + \Gamma(H))(-i\omega + \Gamma(H))} \chi_1 (H).
\end{equation}


\noindent
Here 2$\chi_2(H) = \partial^2 M/\partial H^2, \chi_1 (H) = \partial M/\partial H, M$ is the magnetization of the single-domain region and $\Gamma(H)$ is its relaxation rate. The magnetization of the domain is given by
\begin{equation}                               
 M(mH/T, KN/T) = mL(H/C_1, \alpha),   
\end{equation}


\noindent
 where  $m = \mu N, \alpha = KN/T, C_1= T/m$  and 
\begin{equation}                               
L(H/C_1, \alpha) = C_1 \partial \ln Z/\partial H, \qquad  Z = \int \limits_{0}^{1}\coth (xH/C_1) \exp (\alpha x^2)dx.
\end{equation}


\noindent
Here Z is the partition function. In the first approximation, we perform a fit of the data by neglecting the second net dynamic term in Eq.~(3) and ignoring an $H$-dependence of $\Gamma$. The $M_2$ of the sample, which is the sum of the response from all the F clusters $\widetilde {M}_{2S} = \widetilde C \widetilde {M}_2$  and a contribution from the paramagnetic matrix $M_{2C}$, can be written as 
\begin{equation}                               
M_2(H, \omega)/h^2 = (C'_2 + iC''_2) \chi_2 (H)/m + (C'_3 + iC''_3)H,   
\end{equation}


\noindent
with
\begin{eqnarray}
2C'_2 = \frac{1}{3}\mu\widetilde C \frac{1}{(2\omega/\Gamma)^2 + 1}, \qquad  C''_2 = \frac{2\omega}{\Gamma} C'_2, \\
6C'_3 = (1 - \widetilde C)\frac{\partial^2 M_C}{\partial H^2}\bigg|_{H=0}, \qquad  C''_3 = \frac{2\omega}{\Gamma_C} C'_3,  \qquad \mbox{at} \quad  \Gamma_C \gg 2\omega.  
\end{eqnarray}


\noindent
Here a factor (1/3) in Eq.~(7) takes into account the twinning structure of the crystal, $M_C$ is the magnetization of the paramagnetic matrix in a weak field when $M_{2C} \propto H$ [4], $\Gamma_C$ is the relaxation rate of the uniform magnetization of the paramagnetic matrix. The factor $C'_3$  corresponds to Re$M_{2C}$ at $\omega$ = 0 because $(2\omega/\Gamma_C)^2 \approx 7.8\cdot 10^{-4}$ at $T \approx$ 143 K [6]. One can perform a fit of the Re$M_2$ and Im$M_2$ component of the signal using Eqs.~(4)-(8) where $C_1, C'_2, C''_3, C'_3, C''_3$ and $\alpha$ are the fitting parameters. It allows one to find the following characteristics of the F regions: $\widetilde C_{fit} = \widetilde C<S>, N_{fit} = N<S>, KN$ and $\Gamma$ which are determined by $C'_2, C_1, \alpha$ and $C''_2/C'_2$, respectively.  The value of $<S>$ can be found by exploiting the data on the neutron depolarization (see below). As a result, we obtain all the needed parameters in this approach. A combined fit of the Re$M_2$, Im$M_2$ and neutron depolarization data gives $C_1 \approx 42, \alpha \approx 3, N \approx 4.23\cdot 10^4, \widetilde C \approx 6.75\cdot 10^{- 5}, <S> \approx$ 0.595 as well as $\Gamma/\omega \approx$ 5. We have ignored above an $H$-dependence of $\Gamma$. Now we consider an effect of $\Gamma (H)$. At further evaluation a variation of $\alpha$ makes the fit worse and we exploit below the found above $\alpha$  = 3 value. Using the numerical data for $\Gamma (H)$ which is governed by thermoactivation over the barrier related to the magnetic anisotropy [19], one can obtain an approximate expression at $\alpha$  = 3 
\begin{equation}                               
2\Gamma (H) = \Gamma_0 \frac{\exp [-3(1 + \widetilde H)^2] + \exp [-3(1 - \widetilde H)^2] }{1 + 0.25\widetilde H^2 + 1.5 \widetilde H^2 (1 -  \widetilde H^2)},
\end{equation}


\noindent
where  $ \widetilde H = H/2H_A, H_A = K/\mu$  is the anisotropy field, $ \widetilde H \leq 1, \; \partial \Gamma(H)/\partial H  > 0$ and $\Gamma(H)/\Gamma(0) \approx 8$. At a following fit we use expression (3) for Re$\widetilde M_2$ and Im$\widetilde M_2$  with $\Gamma(H)$ (Eq. (9)) in Eq. (6) instead of the first term. The fitting parameters are $\widetilde C_{fit} = \widetilde C<S>, C_1, \Gamma_0$ and $C'_3$. The $M_2$ data give $\widetilde C_{fit} = \widetilde C<S>$ and $N_{fit} = N<S>$. As a result, one can find the $<S>$ from Eq.(1) assuming a spherical form of the regions $(R = ((3/4\pi)NV_0)^{1/3} \propto (N_{fit} /<S>)^{1/3} \propto 1/<S>^{1/3}, \widetilde C^{1/3} = (\widetilde C_{fit}/<S>)^{1/3} \propto 1/<S>^{1/3}, B^2 R\widetilde C^{1/3} \propto <S>^{4/3})$.

Let us consider the experimental results. Fig.~2 presents the $M_2(H)$ dependences with the fitting curves ($h \approx$ 36.6 Oe). The parameters of the F regions are found to be $C_1 \approx$ 47 Oe, $ N \approx 3.26\cdot 10^4, <S> \approx 0.59, R \approx$ 77.4 \AA, $\widetilde C \approx 8.3 \cdot 10^{-5}, H_A \approx 165$ Oe, $K \approx 1.31\cdot 10^{-2}$ K and $\Gamma(0)/\omega = \Gamma_0 e^{-3}/\omega \approx 6.33, \Gamma \approx$ 624 MHz. We used $P_0 \approx$ 0.938 and $P \approx$ 0.93 at 142.6 K (Fig.~1). Note that Re$\widetilde M_2(H)$ (Eq.~(3)) is determined mainly by the static amplitude $\chi_2(H)$ because: (i) $(2\omega/\Gamma(0))^2 \approx$ 0.1 and $\Gamma(H)$ increases with increasing $H$; (ii) as the calculations show, the contribution of the second dynamical term in Eq.~(3) to Re$M_2$ contains a small factor  $[\omega/\Gamma(H)]^2 < 0.02$ in comparison with the first one. At the same time, the imaginary parts of both terms have the comparable magnitudes and the similar $H$-dependences. We have neglected above a contribution of the twins with the anisotropy axes directed perpendicular to the field which is characterized by $\chi_{2\perp}(H)$. A ratio $r(H) = 2\chi_{2\perp}(H)/\chi_2(H)$ increases with increasing $H$, remaining, however, small in an essential region of changing $\chi_2(H): r(H) < 0.1$ from $H$ = 0 up to a maximum of $|\chi_2(H)|$ at $H^{\mbox{{\scriptsize Re}}}_m \approx$ 40 Oe and $r(H) \approx1$ only at 120 Oe where $|\chi_{2\perp}(H)|$ reaches a maximum. However, at this field $|\chi_2(H)|$ becomes small $\chi_2(120 \mbox{Oe})/\chi_2(H^{\mbox{\scriptsize Re}}_m) \approx$ 0.15 and Re$M_2(H)$ is determined here  by the large linear critical contribution. As a result, the including of $|\chi_{2\perp}(H)|$ does not affect the parameters of the F regions. A frequency dependence of $\chi_{2\perp}(\omega,H)$ was not analyzed. It is known a linear susceptibility $\chi_{1\perp}(\omega)$ at $H$ = 0 whose $\omega$-dependence is determined by a temperature spread of the resonance frequencies of the magnetic moments in the field of anisotropy [20] so that Im$\chi_{1\perp}(\omega) \sim (\omega/2K)\chi_{1\perp}$ for $\omega < 2K \; (\approx$ 613 MHz). One can expect that in the magnetic field Im$\chi_{2\perp}(\omega,H) \sim (\omega/K)\chi_{2\perp}(H)$ at the small $\omega$ and $H < 2H_A \; (\approx$ 330 Oe). Since Im$\chi_2(\omega, H ) \sim 2(\omega/\Gamma(H))\chi_2(H)$ and $\Gamma(H) < 2K$ at $H < 120$ Oe we find, as for Re$\widetilde {M}_2(\omega,H)$, that the contribution of the ``orthogonal'' twins can only affect the tail of Im$\widetilde {M}_2(\omega,H)$ at $H > 120$ Oe where $r(H) > 1$. These corrections are unessential because the parameters of the F domains are mainly determined by a region of $H < 120$ Oe and the linear critical paramagnetic contribution  starts to dominate above 120 Oe. Note, the taking into account the $\chi_{2\perp}(\omega,H)$ contribution to the response could probably improve the fit of the tails of the experimental curves but this question is out the scope of this paper. 

Let us discuss the results. The found attempt frequency $\Gamma_0 \approx 1.25\cdot 10^4$ MHz is within the typical interval of the values ($10^3 - 10^5$ MHz). The magnetization of the F regions characterized by $<S>$ is small in comparison with a saturation magnetization ($<S>/S \approx 0.3, \; S = 1.875$ is the average spin value for the given composition which corresponds to the saturation magnetization). It seems to be the natural result for the temperature slightly below $T\mbox{*} \approx$ 146 K. In the La$_{0.9}$Sr$_{0.1}$MnO$_3$ manganite, the magnetization of the F clusters increased with decreasing temperature and was also small just below $T$* [10]. In regard to the magnetic anisotropy, it is known only an upper limit of the single ion magnetocrystalline anisotropy in the bulk ($J_A S^2_{\alpha}$  with $J_A < 0.5$ K) from the ESR measurements above $T_C$ [7]. For the F regions, $K$ can be written as $K = J_A<S>^2$, leading to $J_A \approx 0.04$ K in agreement with the estimation. In addition, according to our preliminarily ESR data on the $x$ = 0.3 NdBa single crystal, the anisotropy field of the F regions is $H_A \sim 250$ Oe which is not far from $H_A \approx 165$ Oe in our compound. The anisotropy fields of the F clusters were reported for La$_{0.9}$Sr$_{0.1}$MnO$_3$ [10] and La$_{0.88}$Ba$_{0.12}$MnO$_3$ [21]. The dominating uniaxial components, which correspond to the hard and easy axes anisotropies in the first and second manganite, respectively, were found to be about 2.4 kOe. The small value of $H_A \propto  J_A<S>$  in our case ($H_A/(2.4$ kOe) $\approx 0.07$) can be explained by a smallness of $<S>$ and $J_A$. The presented data on the anisotropy fields [10, 21] were obtained at temperatures of 50 K below $T$* that suggested a larger value of $<S>$ (2-3 times) than that at temperature of the several degrees below $T$*. The constant $J_A$ is related to the distortions of the oxygen octahedron and decreases with doping increasing. This constant can be essentially smaller (7-5 times) near the border of the insulator to metal transition in our system than that in the two above mentioned manganites where doping is slightly above a border of an antiferromagnet to ferromagnet transition. 

We obtain the rather large value for the size of the F regions $R/a \approx$ 19.9, where $a = (V_0)^{1/3} \approx$ 3.884 \AA \, is the magnetic lattice constant. It is essential, therefore, to compare it with a critical radius of a single F domain $R_C/a \approx 2\cdot (3J/\omega_D)^{1/2} \; (J$ is the exchange interaction and $\omega_D = 4\pi (g\mu_B)^2/V_0$ is the dipolar energy) for a weakly anisotropic ferromagnet when $2J_A/\omega_D \ll 1 \; (\omega_D \approx$ 0.75 K and  $2J_A/\omega_D \approx$ 0.035 in our system) [22]. Using the mean - field expression for $T_C \; (\approx T\mbox{*} \approx$ 150 K) of the F regions, we find $J \approx$ 14 K and $R_C/a \approx$ 15. This result suggests that the F regions are close to the uniformly magnetized domains. 

As it was discussed in Refs.~(4,6), temperature evolution of the signal of the F regions corresponded mainly to an increase in their volume at decreasing temperature from $T$* down to $T_C$. The matter is that a field position of the extreme ($H^{\mbox{\scriptsize Im}}_m$ 30 Oe, Fig. 2b) in Im$M_2(T,H)$, which is due to the F regions, remains temperature independent with decreasing temperature whereas the extreme value of Im$M_2(T,H^{\mbox{\scriptsize Im}}_m)$ increases sharply down to $T_C$. This behavior reflects an increase in the concentration of the F clusters without the noticeable changes in their parameters. The same peculiarity of the Im$M2(T,H^{\mbox{\scriptsize Im}}_m)$ dependence with $H^{\mbox{\scriptsize Im}}_m \approx$ 22 Oe was also observed in the $x$ = 0.23 NdBa crystal [4]. The closeness of $H^{\mbox{\scriptsize Im}}_m$ and a strong resemblance between the Re$M_2(T,H)$ functions in both systems suggests that the F regions are characterized by the close values of the parameters. The main difference between the compounds is the concentration of the F clusters which is about 10 times larger at $x$ = 0.25 than that at $x$ = 0.23 [6]. 

Magnetization of the F regions is characterized by the small scale of the field nonlinearity $C_1 \approx$ 47 Oe (Eq.~4). As a result, the $M(T,H)$ measurements at $H$ = 1 kOe performed above $T_C$ in the $x$ = 0.25 NdBa compound reveal only the critical temperature dependence of the paramagnetic matrix [7] since the magnetization of the F clusters, being near to a saturation for the superparamagnetic regime, gives a temperature independent correction which is proportional to their rather small concentration. An extreme sensitivity of the clustered state to magnetic field was found in La$_{1-x}$Ba$_x$MnO$_3 \; (x$ = 0.27 and 0.3) [13]. The anomalous contribution to $\chi_1$ was suppressed by $H \sim$ 150 Oe in the paramagnetic region (for $T$ = 275-350 K, $T_C \approx$ 245 K, $x = 0.27$ and $T$ = 310-345 K, $T_C \approx$ 310 K, $x = 0.3$). This behavior can be explained by a small characteristic $H$-scale for magnetization of the F clusters which has the comparable value with one in our system. 

In conclusion, the analysis of the data on the nonlinear longitudinal response and the depolarization of a polarized neutron beam for the Nd$_{0.75}$Ba$_{0.25}$MnO$_3$ manganite allows us to estimate the characteristic size of the F clusters ($R \approx$ 77.4 \AA), which form in critical paramagnetic regime below $T\mbox{*} \approx$ 147 K and lead to a formation of heterogeneous magnetic state above $T_C$. 

This work was supported by the Program No 27 of the PRAS and by the RFBR (grant No 09-02-01509).


\newpage
{\bf Figure captions}

\noindent
Fig.~1. Temperature dependence of the depolarization of the polarized neutron beam for Nd$_{1-x}$Ba$_x$MnO$_3$ ($x$ = 0.23, 0.25) powdered samples. Neutron wavelength is $\lambda$ = 9.2 \AA, $\triangle \lambda/\lambda$ = 0.25 and the thickness of the sample is $L \approx$ 0.2 cm. 

\noindent
Fig.~2. Second harmonic of magnetization, $M_2(H)$, as a function of $H$ for the Nd$_{1-x}$Ba$_x$MnO$_3$ single crystal. Panels show $H$-dependence of the phase components ((a) - Re$M_2(H)$, (b) - Im$M_2(H)$) at $T$ = 142.6 K and their fits. The fits are described in the text.

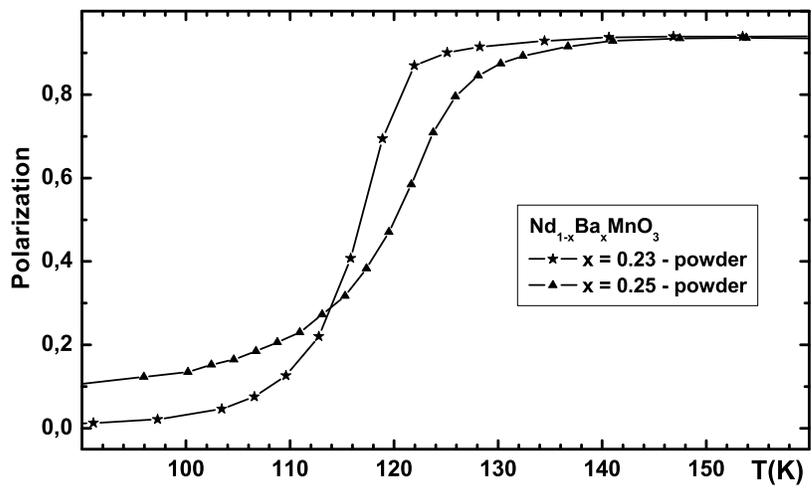
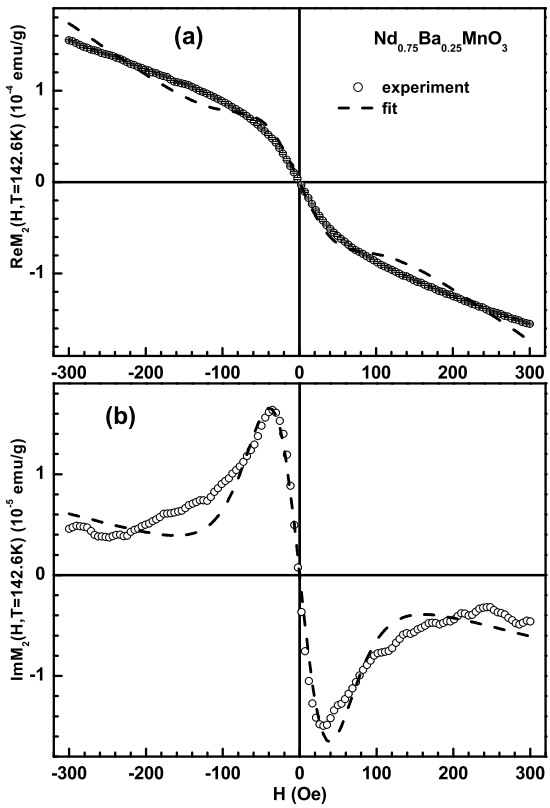
\end{document}